\documentclass[conference]{IEEEtran}
\IEEEoverridecommandlockouts
\usepackage{cite}
\usepackage{amsmath,amssymb,amsfonts}
\usepackage{algorithmic}
\usepackage{graphicx}
\usepackage{textcomp}
\usepackage[table,xcdraw]{xcolor}
\def\BibTeX{{\rm B\kern-.05em{\sc i\kern-.025em b}\kern-.08em
    T\kern-.1667em\lower.7ex\hbox{E}\kern-.125emX}}
\begin{document}

\title{Anomaly Detection in Medical Imaging - A Mini Review}
\author{\IEEEauthorblockN{Maximilian E. Tschuchnig\IEEEauthorrefmark{1}\IEEEauthorrefmark{2} and Michael Gadermayr\IEEEauthorrefmark{1}}
\IEEEauthorblockA{\IEEEauthorrefmark{1}Information Technology and Systems Management \\
Salzburg University of Applied Sciences\\
\IEEEauthorrefmark{2}Artificial Intelligence and Human Interfaces\\
University of Salzburg}}

\maketitle

\begin{abstract}
The increasing digitization of medical imaging enables machine learning based improvements in detecting, visualizing and segmenting lesions, easing the workload for medical experts. However, supervised machine learning requires reliable labelled data, which is is often difficult or impossible to collect or at least time consuming and thereby costly. Therefore methods requiring only partly labeled data (semi-supervised) or no labeling at all (unsupervised methods) have been applied more regularly. Anomaly detection is one possible methodology that is able to leverage semi-supervised and unsupervised methods to handle medical imaging tasks like classification and segmentation. This paper uses a semi-exhaustive literature review of relevant anomaly detection papers in medical imaging to cluster into applications, highlight important results, establish lessons learned and give further advice on how to approach anomaly detection in medical imaging. The qualitative analysis is based on google scholar and 4 different search terms, resulting in 120 different analysed papers. The main results showed that the current research is mostly motivated by reducing the need for labelled data. Also, the successful and substantial amount of research in the brain MRI domain shows the potential for applications in further domains like OCT and chest X-ray.
\end{abstract}


\section{Introduction}
The increasing digitization of medical imaging enables the collection of data and machine learning (ML) based approaches to aid medical experts. One powerful part of ML comes from supervised methods, using both data and corresponding labels in e.g. segmentation or classification models. However, since the collection of annotations (labels) is often times time consuming and thereby costly~\cite{domingues2018comparative} as well as in many cases a confident ground truth even being unobtainable, their usability is reduced. Due to this, semi-supervised and unsupervised methods are applied. This is often achieved through anomaly detection.

\textit{Definitions: }Pathologies in medical images can often be described as a rare deviance from a norm, or a non-anomalous (in the case of medical imaging mostly healthy) sample. This fits the definition of outliers (or anomalies) in the data, motivating the application of anomaly detection~\cite{grubbs1969procedures}. In this publication, the terms anomaly detection and outlier detection are used interchangeably. This is motivated by the fact that outliers are sometimes defined as valid but out of order datapoints, while anomalies also include further differences (e.g. different image capture modalities). Therefore outliers can be defined as a subset of anomalies. Anomaly detection can be separated into $3$ classes, \textit{point}, \textit{collective} and \textit{contextual} anomalies. Point anomaly detection is the task of recognizing a single anomalous point from a larger dataset. Most anomaly detection models handle point anomalies. Collective anomalies are anomalies that may not be identified as anomalies if viewed as a single point but as a set of many they form an anomaly. Contextual anomalies can only be recognized as anomalies if context is added. There are also $3$ different anomaly detection setups, \textit{supervised}, \textit{semi-supervised} and \textit{unsupervised} anomaly detection. Supervised anomaly detection is comparable with classification using a very unbalanced dataset. Semi-supervised anomaly detection aims to train a model on only one, typically the normal (in our case healthy) class and then applies the model to both healthy and pathological data, reporting the corresponding scores. Unsupervised anomaly detection uses both, normal and anomalous data, does not make use of labels at all and works purely on intrinsic properties of the dataset (using distances or densities) \cite{goldstein2016comparative}. In anomaly detection, the usage of semi-supervised and unsupervised anomaly detection (UAD) is confused, and repeatedly applied to both semi-supervised and unsupervised methods. We believe that the separation into semi-supervised (healthy data being clearly defined) and unsupervised (no definition of labels at all) makes sense and advise to use this terminology as also pointed out by~\cite{goldstein2016comparative}.

\textit{Deviation based anomaly detection: }
Anomaly detection using medical image data, e.g. computed tomography (CT) scans, is typically performed using either convolutional neural network (CNN) based feature extractors, followed by one-class (OC) classifiers or deviation based methods like autoencoders (AEs)~\cite{myvincent2010, mysun2018, myuzunova2019} or even more recently, generative adversarial network (GAN)~\cite{schlegl2017unsupervised, schlegl2019f ,baur2018deep} based methods. Both AEs and GANs use convolutional kernels, however their applications in the sense of deviation based anomaly detection are fundamentally different to CNN based feature extractors. In order to generate deviation based scores from an AE, the encoder of the encoder-decoder based neural network typically encodes a sample image into a lower dimensional latent space, also called a bottleneck. The decoder uses this latent space representation to recreate the sample and a deviation between the sample and the reconstruction can be calculated. During training, this deviation is used to backpropagate and update the network. The AE in an anomaly detection setting is trained using healthy data to en- and decode features of healthy samples, leading to a higher deviation for non-healthy samples, assuming that there is a difference between the learned healthy and the lesioned latent space~\cite{mygong2019}. 
GANs can also be used to facilitate a deviation based score. In addition to training a generator and a discriminator in an adversarial setup, an additional encoder needs to be trained, mapping the generated image back to the latent space (input to generator)~\cite{schlegl2017unsupervised}. By doing this, any input image can be mapped to a latent space and reconstructed into an image using the generator. This results in a reconstruction which can be used to facilitate a reconstruction loss. 

Additionally there are conventional methods to facilitate anomaly detection, using e.g. z-score thresholds~\cite{li2015outlier, li2015burn}, boxplots~\cite{li2016quality} or methods built on the ideas of principal components analysis (PCA)~\cite{mejia2017pca, kim2021chest}. OC support vector machines (SVM)s \cite{tax2001uniform} are one of the most known semi-supervised anomaly detection methods. In principle they apply the ideas of SVMs (using hyperplanes to separate two classes using support vectors with the aim of generating the largest possible margin) to a OC problem. One possibility to achieve this is to model a hypersphere to encompass all support vectors, creating the smallest possible hypersphere.  

\textit{Contribution:} This papers contribution is the analysis of the current state of anomaly detection in medical imaging. Using this analysis, we show lessons learned and give an outlook for future applications and research targets.

\section{Method}
The method used was a semi-exhaustive literature review based on Randolph~\cite{randolph2009}. The formulated problem was the evaluation of anomaly detection in medical imaging. For data collection, the search engine Google Scholar was used. In order to obtain meaningful results, the search terms \textit{anomaly detection in medical imaging}, \textit{unsupervised anomaly detection in medical imaging}, \textit{outlier detection in medical imaging} and \textit{unsupervised outlier detection in medical imaging} were chosen. From these results the following criteria for exclusion were chosen. Only the first $3$ pages of results (sorted by relevance, $10$ articles per page) were used. Further, the criteria for exclusion \textit{duplicate}, \textit{in context of medical imaging (in abstract, title or conclusion)}, \textit{peer-review} and \textit{date} were identified. Since the search terms were similar, \textit{duplicates} had to be removed. Papers without a clear focus on \textit{medical imaging} in the abstract, title or conclusion were also removed. A further criterion was to only include \textit{peer-reviewed} research items. This mainly lead to the exclusion of preprints. The data timeframe was set to not include papers after the resurgence of deep-learning (AlexNet~\cite{krizhevsky2012}) and to still include papers after the U-net was proposed~\cite{ronneberger2015}, resulting in a timeframe of January $2015$ $-$ July $2021$. This lead to a reduction of papers from $120$ to $49$. Since these papers also included $4$ survey papers, the final number of application based research papers was $45$. These survey papers were used as a qualitative comparision to the our extracted lessons learned. Next, the papers were manually clustered with respect to their imaging method and the following information was extracted: \textit{Aim}, \textit{Applied Method} and  \textit{Results}. From these clusters, lessons learned were extracted, which are reported in section \ref{sec:results}.

\section{Results}
\label{sec:results}
The semi-exhaustive literature review resulted in $45$ research items, from which further $6$ were removed due to not containing applications in medical imaging (only exemplar stated in abstract) or being non-available. The resulting papers were further clustered into $5$ categories (corresponding to Tab. \ref{tab:ocular}-\ref{tab:mix} by their imaging methods. Tab.~\ref{tab:ocular} shows papers applying anomaly detection to occular medical images with retinal fundus images and optical coherence tomography (OCT).  Tab.~\ref{tab:center} focuses on papers with applications in the center body region, with chest X-rays and mammography. Tab.~\ref{tab:CTfMRI} summarizes application papers, using CT and functional magnetic resonance imaging (fMRI). Tab. \ref{tab:brain} displays papers applying ML to brain Magnetic resonance imaging (MRI) datasets. Tab. \ref{tab:mix} shows mixed applications from the domains of breast ultrasound, chest radiographs, histology and fundus images as well as multi-spectral imaging (MSI).

Overall, these tables show a narrow field of application with $15$ ($38.46\%$) of all selected papers working on MRI scans of the brain. Further $6$ papers use fMRI and CT scans of the brain, increasing the amount of brain image data based applications to $53.85\%$. Further clusters could be observed using chest X-rays and mammography, as well as ocular imaging techniques, especially OCT. Of note is, that although medical imaging includes methods like histology, only $1$ paper~\cite{vstepec2020image} applied anomaly detection to such data. A further result is the relevance of deviation based methods, with $27$ papers ($69.23\%$) applying some form or adaptation, mostly using autoencoders AEs or GANs~\cite{sato2018primitive, fujioka2020efficient, nakao2021unsupervised, vstepec2020image, venkataramanan2020attention, schlegl2017unsupervised, schlegl2019f, seebock2019exploiting, zhao2021anomaly, wolleb2020descargan, tlusty2018unsupervised, wei2018anomaly, pawlowski2018unsupervised, armanious2019unsupervised, khosla2019detecting, atlason2019unsupervised, alaverdyan2018unsupervised, chen2019unsupervised, alaverdyan2020regularized, baur2021modeling, heer2021ood, baur2020steganomaly, baur2018deep, zimmerer2019unsupervised, van2021anomaly, han2021madgan, you2019unsupervised}. Investigating MRIs, $7$~\cite{zuo2018automatic, popescu2021distributional, baur2021modeling, baur2020steganomaly, zimmerer2019unsupervised, han2021madgan, you2019unsupervised} of the $15$ publications using brain MRI data focus explicitly on tumours or metastases, showing the usefulness of anomaly detection and segmentation of tumours in brains using MRI. Most other brain MRI based methods more generally handle the task of lesion classification or segmentation with only two focusing specificly on cerebral small vessel diseases~\cite{bowles2017brain, van2021anomaly}. A further cluster uses X-ray for the detection of pneumonia \cite{nakao2021unsupervised, zhang2020viral} or lung disease like COVID-19~\cite{kim2021chest}. Several advancements have also been made in OCT segmentation of retina lesions~\cite{ouardini2019towards, schlegl2017unsupervised, schlegl2019f, seebock2019exploiting, zhao2021anomaly}, with one publication performing visual touring test using $2$ experts, which were unable to recognize differences in the correctly reconstructed data~\cite{schlegl2019f}. Breast cancer and pathology detection was also improved using anomaly detection \cite{wei2018anomaly, quellec2016multiple, tlusty2018unsupervised}.

\begin{table*}[t]
\centering
\caption{Table consisting of ocular image based results obtained by the literature review}
\label{tab:ocular}
\begin{tabular}{m{1cm}m{2.5cm}m{4cm}m{7cm}}
\textbf{Paper}  & \textbf{Imaging Method}    & \textbf{Aim}                                                        & \textbf{Applied Method}                                                    \\
\cite{ouardini2019towards}  & retinal fundus images & transfer learning (general and retinal   lesions) & TL (IMNet feature extrator)                                           \\
\cite{schlegl2017unsupervised}  & OCT                   & new anomaly detection method                      & AnoGAN                                                                \\
\cite{schlegl2019f}  & OCT                   & new anomaly detection method                      & fAno-GAN                                                              \\
\cite{seebock2019exploiting} & OCT                   & segmentation (retina lesions)                     & Bayesian U-Net. Episdemic uncertainty estimations and post processing \\
\cite{zhao2021anomaly} & OCT and chest X-ray   & new anomaly detection method                      & encoder-decoder with additional GAN discriminators                   
\end{tabular}
\end{table*}

\begin{table*}[t]
\centering
\caption{Table consisting of center body image based results obtained by the literature review}
\label{tab:center}
\begin{tabular}{m{1cm}m{2.5cm}m{4cm}m{7cm}}
\textbf{Paper}  & \textbf{Imaging Method}    & \textbf{Aim}                                                        & \textbf{Applied Method}          \\
\cite{nakao2021unsupervised} & chest radiographs & anomaly detection (pneumonia)   & $\alpha$-GAN   \\
\cite{zhang2020viral} & chest X-ray       & anomaly detection (virial pneumonia)                                & CNN feature extractor with anomaly score (Fully connected) and confidency (Fully connected) \\
\cite{wolleb2020descargan} & chest X-ray       & new anomaly detection method (pleural effusions)                    & DeScarGAN                                                                           \\
\cite{kim2021chest} & chest X-ray       & anomaly detection (coronavirus)                                     & edge detection and morphology. PCA to reduce features and use in RNN               \\
\cite{tlusty2018unsupervised} & mammography & anomaly detection (compressions or implants) & Stacked AE as feature extractor, K-Means for clustering                             \\
\cite{quellec2016multiple} & (MIL) mammography  & anomaly detection (breast cancer)                                   & Simultaniously trained MIL algorithms (DD, APR, and MILBoost)                       \\
\cite{wei2018anomaly} & mammography          & anomaly detection (breast anomalies)                                & cAE with RMSD threshold                                                            
\end{tabular}
\end{table*}

\begin{table*}[t]
\centering
\caption{Table consisting of CT and fMRI image based results of the brain obtained by the literature review}
\label{tab:CTfMRI}
\begin{tabular}{m{1cm}m{2.5cm}m{4cm}m{7cm}}
\textbf{Paper}  & \textbf{Imaging Method}    & \textbf{Aim}       & \textbf{Applied Method}  \\
\cite{sato2018primitive} & head CT (3D)  &anomaly detection (emergency head CTs)  & 3D cAE \\
\cite{pawlowski2018unsupervised} & brain CT (2D)  & anomaly detection (brain   lesions)       & Bayesian AE \\
\cite{armanious2019unsupervised} & PET-CT and brain MRI & image-to-image translation (image artifacts)                          & Cycle-MedGAN                               \\
\cite{mejia2017pca} & Brain fMRI           & pca based outlier removal (image artifacts)                           & PCA (robust distance and leverage)         \\
\cite{khosla2019detecting} & brain rs-fMRI        &                                                                       & AE and frame prediction (conv-LSTM)        \\
\cite{kuo2016framework} & Brain fMRI           & anomaly detection using constraint programming (cognitive impairment) & Constraint Programming using 3 constraints
\end{tabular}
\end{table*}

\begin{table*}[t]
\centering
\caption{Table consisting of fMRI image based results of the brain obtained by the literature review}
\label{tab:brain}
\begin{tabular}{m{1cm}m{2.5cm}m{4cm}m{7cm}}
\textbf{Paper}  & \textbf{Imaging Method}    & \textbf{Aim}                                                        & \textbf{Applied Method}     \\
\cite{atlason2019unsupervised} & brain MRI       & segmentation (brain lesions)                      & SegAE                                                                 \\
\cite{alaverdyan2018unsupervised} & brain MRI       & anomaly detection (epilepsy)                      & siamese network, stacked cAE, wasserstein AE                          \\
\cite{chen2019unsupervised} & brain MRI       & improvements to AE based methods (glioma)         & VAE + LG (and several baselines)                                      \\ 
\cite{bowles2017brain} & brain MRI       & segmentation (cerebral small vessel disease)      & PHI-Syn \cite{bowles2016pseudo} (image synthesis) and Gaussian mixture models used by oc-SVM \\
\cite{zuo2018automatic} & Brain MRI       & segmentation (brain lesions)                      & Hidden markov models                                                                   \\
\cite{alaverdyan2020regularized} & Brain MRI       & anomaly detection (brain lesion)                  & siamese, stacked cAE for latent representations in oc-SVM     \\
\cite{popescu2021distributional} & Brain MRI       & segmentation (brain tumor)                        & DistGP-Seg. Incooperating DistGP into CNN                             \\ 
\cite{baur2021modeling} & Brain MRI       & anomaly detection (MS and cancer)                 & spatial AE with skip connections                                      \\
\cite{heer2021ood} & Brain MRI       & awareness for OOD                                 & VAE. Scores: l1, Kullback–Leibler divergence, Watanabe–Akaike information criterion score, Density of States Estimation                                    \\
\cite{baur2020steganomaly} & Brain MRI       & improvements to cycleGAN (brain tumor)            & SteGANomaly                                                           \\
\cite{baur2018deep} & Brain MRI       & anomaly segmentation (brain lesions)              & AnoVAEGan                                                             \\
\cite{zimmerer2019unsupervised} & Brain MRI       & anomal localization (brain tumor)                 & VAE with additional KL divergence term in Backprop                    \\
\cite{van2021anomaly} & Brain MRI       & anomaly detection (brain infarct)                 & GANomaly                                                              \\
\cite{han2021madgan} & brain MRI       & new anomaly detection method (brain mestastases)  & (Wasserstein based) MaDGAN using self attention (paired)              \\
\cite{li2016quality} & Brain MRI (DTI) & quality assurance of segmentation (brain lesions) & non parametric (box-plots); supervised classification models      \\
\cite{you2019unsupervised} & Brain MRI       & new anomaly detection method (tumor)              & GMVAE                                                                
\end{tabular}
\end{table*}

\begin{table*}[t]
\centering
\caption{Table consisting of remaining mixed literature review results}
\label{tab:mix}
\begin{tabular}{m{1cm}m{2.5cm}m{4cm}m{7cm}}
\textbf{Paper}  & \textbf{Imaging Method}    & \textbf{Aim}                                                        & \textbf{Applied Method}                                         \\
\cite{fujioka2020efficient} & breast ultrasound & anomaly detection (normal, begning, malignant in breasts) & bidirectional GAN  \\
\cite{vstepec2020image} & hisotlogy images  & image synthesis (tumor)                                    & DCGAN \& WGAN                                          \\
\cite{venkataramanan2020attention} & fundus image      & anomly localization (glaucoma)                             & adversarial attention guided VAE                       \\
\cite{li2015outlier} & MSI               & outlier removal to improve burn detection                  & z-score based outlier detection to improve SVM and KNN \\
\cite{li2015burn} & MSI               & outlier removal to improve burn detection                  & z-score based outlier detection to improve SVM and KNN
\end{tabular}
\end{table*}

One result of this analysis is the statement that anomaly detection can be motivated by the lack of available labelled training data, which was stated in $19$ publications. The reported results of these papers proved that these semi- and unsupervised approaches successfully completed their tasks~\cite{wei2018anomaly, schlegl2017unsupervised, schlegl2019f, zhao2021anomaly, quellec2016multiple, van2021anomaly, seebock2019exploiting, fujioka2020efficient, ouardini2019towards, zimmerer2019unsupervised, vstepec2020image, heer2021ood, venkataramanan2020attention, zuo2018automatic, alaverdyan2020regularized, tlusty2018unsupervised, mejia2017pca, armanious2019unsupervised, atlason2019unsupervised}. However, some papers also show semi-supervised methods outperforming fully supervised methods. These outperforming methods are based on classical feature extraction followed by multiple-instance learning (MIL) based models~\cite{quellec2016multiple}, through adaptations to GANs~\cite{wolleb2020descargan} (using skip-connections and weight-sharing subnetworks) and through the adaptation of AEs to the SegAE model~\cite{khosla2019detecting} (using pairs of T1-w, T2-w and FLAIR data for improved anomaly detection). For this improvement in comparison to fully supervised models, Khosla et al.~\cite{khosla2019detecting} reason that fully supervised methods systematically either under or overestimate lesion volumes (when segmenting lesions), while their proposed method was reported to be free of this bias.

Zhang et al. and Kim et al. both show interesting approaches, applying conventional feature extractors (CNN and edge detection) with further OC classifiers (fully connected neural networks and recurrent neural networks). By using these methods both papers reach relatively high scores, but still lower scores then their CT based baselines.

A further finding is the obvious bias in the amount of research items regarding OCT, chest X-Ray, mammography and Brain MRI. An investigation in the used datasets shows a strong dataset and community driven effect. For all of the above mentioned image categories, datasets are publicly available. Further, a community driven effect can be observed, comparing new models against older ones, evaluated on the same dataset.

In addition to medical image based application papers, several authors proposed improvements to the general anomaly detection pipelines. $3$ papers showed an improvement of subsequent methods by removing anomalies from the data or reducing complexity in the data~\cite{li2015outlier, li2015burn,mejia2017pca}. Also, constraint programming is shown successfully by Kuo et al.~\cite{kuo2016framework}. showing further approaches to perform anomaly detection. CycleGAN is also shown to work for transforming images into a space that showed reduced image artefacts~\cite{armanious2019unsupervised}. Heer et al.~\cite{heer2021ood} showed issues with the general idea of anomaly detection and their application of anomalies as out-of-distribution (OOD) data, remarking a blind spot using deviation based methods. They state that denoting anomalies as OOD is dangerous, since non anomalous data from different sensors or image modalities may also be detected as OOD although this data not being anomalous. In their paper they further present a method based on prior knowledge to disentangle lesion based OOD from non-lesion based counterparts.


\section{Discussion}
In this paper we analysed the current state of research in anomaly detection using medical image data and extracted lessons learned. To accomplish this, a semi-exhaustive literature review was performed, resulting in 120 papers, from which 44 were further investigated (after filters were applied). This resulted in $4$ major clusters of image domains, with the brain MRI domain comprising $39.45\%$ of all papers.

One takeway is that especially in the brain MRI domain, both lesion and tumour classification as well as segmentation have been successfully implemented multiple times. It is shown that both AE and GAN based methods as well as Gaussian mixture models, hidden Markov models and CNNs with specific feature extractors can work in this anomaly detection setup. This was further shown to be the case with chest X-ray, mammography as well as OCT data. Extrapolating from these results, first approaches in similar domains, using anomaly detection for tasks in the domains of e.g. CT scans of the skeleton or spines seem promising and should be investigated. Also, an investigation of the suitability for histological data would be of high interest, since histological data was very under-represented ($1$ publication). However, there are multiple differences between CT/MRI and histology. In histology it is not sufficient to detect a large object (e.g. tumor) which is indicated by different intensity values. It would rather be important to learn the shape and interaction of nuclei and cells which is supposed to be a more challenging task, relying more on high frequency information which is a reported weak point of several proposed deviation based mehtods. Further, histology images are extremely high resolution, leading to issues using current GAN or AE based anomaly detection. Štepec et al.~\cite{vstepec2020image} show one way to circumvent these issues successfully using patch extraction and MIL.

As reported in the results, there were some semi-supervised anomaly detection models that resulted in higher or similar scores than their fully supervised alternatives. One interpretation is that, especially regarding segmentation, human labelled segmentation masks with rough edges may introduce bias. This is however still unclear and should further be investigated.

Another useful takeaway is that not only improvements to state of the art (SOTA) models are needed but also simpler models or cheaper image modalities can be a major improvement, even if the SOTA scores cannot be reached e.g. replacing CT with X-ray based methods. One example was shown by Zhang et al.~\cite{zhang2020viral} who used X-ray images, approaching relatively high scores. Although their method did not outperform the CT based baselines, the methods is still of high significance, since it reached similar levels using X-rays requiring a lower radiation dose and an more available imaging method.

As stated by~\cite{baur2021autoencoders} we also recognized the generation of free and comparable datasets as a high priority to facilitate further research. The fast growing brain MRI community showed, that open datasets are an important asset to boost research. Therefore the development and open distribution should be pursued for different medical image domains. In order to facilitate anomaly detection research, a semi-supervised dataset (only including a small amount of annotations) should be developed.

A disadvantage, reported by several deep learning based approaches was~\cite{pawlowski2018unsupervised, zuo2018automatic}, that results were still unstable and more research was needed before a clinical application could be performed. This however was not always the case~\cite{fujioka2020efficient} but there are still doubts in the clinical applicability of deep learning based anomaly detection methods. Large clinical application studies would be needed to show their suitability.

\textit{Conclusion:} 
In this paper we investigated the current state of research in medical image based anomaly detection and generated lessons learned. The lessons learned can be converted into the following future targets: \textit{a very narrow domain of application that should be expanded}, \textit{development of freely accessible datasets}, \textit{investigation of the OCT blindspot} and \textit{improvements of working approaches like constraints on the AE bottleneck}.

\section*{Acknowledgment}
This work was partially funded by the County of Salzburg under grant number FHS-2019-10-KIAMed.

\nocite{*}
\bibliography{conference_idsc}{}

\begin{thebibliography}{10}
\providecommand{\url}[1]{#1}
\csname url@samestyle\endcsname
\providecommand{\newblock}{\relax}
\providecommand{\bibinfo}[2]{#2}
\providecommand{\BIBentrySTDinterwordspacing}{\spaceskip=0pt\relax}
\providecommand{\BIBentryALTinterwordstretchfactor}{4}
\providecommand{\BIBentryALTinterwordspacing}{\spaceskip=\fontdimen2\font plus
\BIBentryALTinterwordstretchfactor\fontdimen3\font minus
  \fontdimen4\font\relax}
\providecommand{\BIBforeignlanguage}[2]{{%
\expandafter\ifx\csname l@#1\endcsname\relax
\typeout{** WARNING: IEEEtran.bst: No hyphenation pattern has been}%
\typeout{** loaded for the language `#1'. Using the pattern for}%
\typeout{** the default language instead.}%
\else
\language=\csname l@#1\endcsname
\fi
#2}}
\providecommand{\BIBdecl}{\relax}
\BIBdecl

\bibitem{domingues2018comparative}
R.~Domingues, M.~Filippone, P.~Michiardi, and J.~Zouaoui, ``A comparative
  evaluation of outlier detection algorithms: Experiments and analyses,''
  \emph{Pattern Recognition}, vol.~74, pp. 406--421, 2018.

\bibitem{grubbs1969procedures}
F.~E. Grubbs, ``Procedures for detecting outlying observations in samples,''
  \emph{Technometrics}, vol.~11, no.~1, pp. 1--21, 1969.

\bibitem{goldstein2016comparative}
M.~Goldstein and S.~Uchida, ``A comparative evaluation of unsupervised anomaly
  detection algorithms for multivariate data,'' \emph{PloS one}, vol.~11,
  no.~4, p. e0152173, 2016.

\bibitem{myvincent2010}
P.~Vincent, H.~Larochelle, I.~Lajoie, Y.~Bengio, P.-A. Manzagol, and L.~Bottou,
  ``Stacked denoising autoencoders: Learning useful representations in a deep
  network with a local denoising criterion.'' \emph{Journal of machine learning
  research}, vol.~11, no.~12, 2010.

\bibitem{mysun2018}
J.~Sun, X.~Wang, N.~Xiong, and J.~Shao, ``Learning sparse representation with
  variational auto-encoder for anomaly detection,'' pp. 33\,353--33\,361, 2018.

\bibitem{myuzunova2019}
H.~Uzunova, S.~Schultz, H.~Handels, and J.~Ehrhardt, ``Unsupervised pathology
  detection in medical images using conditional variational autoencoders,''
  \emph{International journal of computer assisted radiology and surgery},
  vol.~14, no.~3, pp. 451--461, 2019.

\bibitem{schlegl2017unsupervised}
T.~Schlegl, P.~Seeb{\"o}ck, S.~M. Waldstein, U.~Schmidt-Erfurth, and G.~Langs,
  ``Unsupervised anomaly detection with generative adversarial networks to
  guide marker discovery,'' in \emph{International conference on information
  processing in medical imaging}.\hskip 1em plus 0.5em minus 0.4em\relax
  Springer, 2017, pp. 146--157.

\bibitem{schlegl2019f}
T.~Schlegl, P.~Seeb{\"o}ck, S.~M. Waldstein, G.~Langs, and U.~Schmidt-Erfurth,
  ``f-anogan: Fast unsupervised anomaly detection with generative adversarial
  networks,'' \emph{Medical image analysis}, vol.~54, pp. 30--44, 2019.

\bibitem{baur2018deep}
C.~Baur, B.~Wiestler, S.~Albarqouni, and N.~Navab, ``Deep autoencoding models
  for unsupervised anomaly segmentation in brain mr images,'' in
  \emph{International MICCAI Brainlesion Workshop}.\hskip 1em plus 0.5em minus
  0.4em\relax Springer, 2018, pp. 161--169.

\bibitem{mygong2019}
D.~Gong, L.~Liu, V.~Le, B.~Saha, M.~R. Mansour, S.~Venkatesh, and A.~v.~d.
  Hengel, ``Memorizing normality to detect anomaly: Memory-augmented deep
  autoencoder for unsupervised anomaly detection,'' in \emph{Proceedings of the
  IEEE/CVF International Conference on Computer Vision}, 2019, pp. 1705--1714.

\bibitem{li2015outlier}
W.~Li, W.~Mo, X.~Zhang, J.~J. Squiers, Y.~Lu, E.~W. Sellke, W.~Fan, J.~M.
  DiMaio, and J.~E. Thatcher, ``Outlier detection and removal improves accuracy
  of machine learning approach to multispectral burn diagnostic imaging,''
  \emph{Journal of biomedical optics}, vol.~20, no.~12, p. 121305, 2015.

\bibitem{li2015burn}
W.~Li, W.~Mo, X.~Zhang, Y.~Lu, J.~J. Squiers, E.~W. Sellke, W.~Fan, J.~M.
  DiMaio, and J.~E. Thatcher, ``Burn injury diagnostic imaging device's
  accuracy improved by outlier detection and removal,'' in \emph{Algorithms and
  Technologies for Multispectral, Hyperspectral, and Ultraspectral Imagery
  XXI}, vol. 9472.\hskip 1em plus 0.5em minus 0.4em\relax International Society
  for Optics and Photonics, 2015, p. 947206.

\bibitem{li2016quality}
K.~Li, C.~Ye, Z.~Yang, A.~Carass, S.~H. Ying, and J.~L. Prince, ``Quality
  assurance using outlier detection on an automatic segmentation method for the
  cerebellar peduncles,'' in \emph{Medical Imaging 2016: Image Processing},
  vol. 9784.\hskip 1em plus 0.5em minus 0.4em\relax International Society for
  Optics and Photonics, 2016, p. 97841H.

\bibitem{mejia2017pca}
A.~F. Mejia, M.~B. Nebel, A.~Eloyan, B.~Caffo, and M.~A. Lindquist, ``Pca
  leverage: outlier detection for high-dimensional functional magnetic
  resonance imaging data,'' \emph{Biostatistics}, vol.~18, no.~3, pp. 521--536,
  2017.

\bibitem{kim2021chest}
C.-M. Kim, E.~J. Hong, and R.~C. Park, ``Chest x-ray outlier detection model
  using dimension reduction and edge detection,'' \emph{IEEE Access}, 2021.

\bibitem{tax2001uniform}
D.~M. Tax and R.~P. Duin, ``Uniform object generation for optimizing one-class
  classifiers,'' \emph{Journal of machine learning research}, vol.~2, no. Dec,
  pp. 155--173, 2001.

\bibitem{randolph2009}
J.~Randolph, ``A guide to writing the dissertation literature review,''
  \emph{Practical Assessment, Research, and Evaluation}, vol.~14, no.~1, p.~13,
  2009.

\bibitem{krizhevsky2012}
A.~Krizhevsky, I.~Sutskever, and G.~Hinton, ``2012 alexnet,'' pp. 1--9, 2012.

\bibitem{ronneberger2015}
O.~Ronneberger, P.~Fischer, and T.~Brox, ``U-net: Convolutional networks for
  biomedical image segmentation,'' in \emph{International Conference on Medical
  image computing and computer-assisted intervention}.\hskip 1em plus 0.5em
  minus 0.4em\relax Springer, 2015, pp. 234--241.

\bibitem{vstepec2020image}
D.~{\v{S}}tepec and D.~Sko{\v{c}}aj, ``Image synthesis as a pretext for
  unsupervised histopathological diagnosis,'' in \emph{International Workshop
  on Simulation and Synthesis in Medical Imaging}.\hskip 1em plus 0.5em minus
  0.4em\relax Springer, 2020, pp. 174--183.

\bibitem{sato2018primitive}
D.~Sato, S.~Hanaoka, Y.~Nomura, T.~Takenaga, S.~Miki, T.~Yoshikawa, N.~Hayashi,
  and O.~Abe, ``A primitive study on unsupervised anomaly detection with an
  autoencoder in emergency head ct volumes,'' in \emph{Medical Imaging 2018:
  Computer-Aided Diagnosis}, vol. 10575.\hskip 1em plus 0.5em minus 0.4em\relax
  International Society for Optics and Photonics, 2018, p. 105751P.

\bibitem{fujioka2020efficient}
T.~Fujioka, K.~Kubota, M.~Mori, Y.~Kikuchi, L.~Katsuta, M.~Kimura, E.~Yamaga,
  M.~Adachi, G.~Oda, T.~Nakagawa \emph{et~al.}, ``Efficient anomaly detection
  with generative adversarial network for breast ultrasound imaging,''
  \emph{Diagnostics}, vol.~10, no.~7, p. 456, 2020.

\bibitem{nakao2021unsupervised}
T.~Nakao, S.~Hanaoka, Y.~Nomura, M.~Murata, T.~Takenaga, S.~Miki, T.~Watadani,
  T.~Yoshikawa, N.~Hayashi, and O.~Abe, ``Unsupervised deep anomaly detection
  in chest radiographs,'' \emph{Journal of Digital Imaging}, pp. 1--10, 2021.

\bibitem{venkataramanan2020attention}
S.~Venkataramanan, K.-C. Peng, R.~V. Singh, and A.~Mahalanobis, ``Attention
  guided anomaly localization in images,'' in \emph{European Conference on
  Computer Vision}.\hskip 1em plus 0.5em minus 0.4em\relax Springer, 2020, pp.
  485--503.

\bibitem{seebock2019exploiting}
P.~Seeb{\"o}ck, J.~I. Orlando, T.~Schlegl, S.~M. Waldstein, H.~Bogunovi{\'c},
  S.~Klimscha, G.~Langs, and U.~Schmidt-Erfurth, ``Exploiting epistemic
  uncertainty of anatomy segmentation for anomaly detection in retinal oct,''
  \emph{IEEE transactions on medical imaging}, vol.~39, no.~1, pp. 87--98,
  2019.

\bibitem{zhao2021anomaly}
H.~Zhao, Y.~Li, N.~He, K.~Ma, L.~Fang, H.~Li, and Y.~Zheng, ``Anomaly detection
  for medical images using self-supervised and translation-consistent
  features,'' \emph{IEEE Transactions on Medical Imaging}, 2021.

\bibitem{wolleb2020descargan}
J.~Wolleb, R.~Sandk{\"u}hler, and P.~C. Cattin, ``Descargan: Disease-specific
  anomaly detection with weak supervision,'' in \emph{International Conference
  on Medical Image Computing and Computer-Assisted Intervention}.\hskip 1em
  plus 0.5em minus 0.4em\relax Springer, 2020, pp. 14--24.

\bibitem{tlusty2018unsupervised}
T.~Tlusty, G.~Amit, and R.~Ben-Ari, ``Unsupervised clustering of mammograms for
  outlier detection and breast density estimation,'' in \emph{2018 24th
  International Conference on Pattern Recognition (ICPR)}.\hskip 1em plus 0.5em
  minus 0.4em\relax IEEE, 2018, pp. 3808--3813.

\bibitem{wei2018anomaly}
Q.~Wei, Y.~Ren, R.~Hou, B.~Shi, J.~Y. Lo, and L.~Carin, ``Anomaly detection for
  medical images based on a one-class classification,'' in \emph{Medical
  Imaging 2018: Computer-Aided Diagnosis}, vol. 10575.\hskip 1em plus 0.5em
  minus 0.4em\relax International Society for Optics and Photonics, 2018, p.
  105751M.

\bibitem{pawlowski2018unsupervised}
N.~Pawlowski, M.~C. Lee, M.~Rajchl, S.~McDonagh, E.~Ferrante, K.~Kamnitsas,
  S.~Cooke, S.~Stevenson, A.~Khetani, T.~Newman \emph{et~al.}, ``Unsupervised
  lesion detection in brain ct using bayesian convolutional autoencoders,''
  2018.

\bibitem{armanious2019unsupervised}
K.~Armanious, C.~Jiang, S.~Abdulatif, T.~K{\"u}stner, S.~Gatidis, and B.~Yang,
  ``Unsupervised medical image translation using cycle-medgan,'' in \emph{2019
  27th European Signal Processing Conference (EUSIPCO)}.\hskip 1em plus 0.5em
  minus 0.4em\relax IEEE, 2019, pp. 1--5.

\bibitem{khosla2019detecting}
M.~Khosla, K.~Jamison, A.~Kuceyeski, and M.~R. Sabuncu, ``Detecting
  abnormalities in resting-state dynamics: An unsupervised learning approach,''
  in \emph{International Workshop on Machine Learning in Medical
  Imaging}.\hskip 1em plus 0.5em minus 0.4em\relax Springer, 2019, pp.
  301--309.

\bibitem{atlason2019unsupervised}
H.~E. Atlason, A.~Love, S.~Sigurdsson, V.~Gudnason, and L.~M. Ellingsen,
  ``Unsupervised brain lesion segmentation from mri using a convolutional
  autoencoder,'' in \emph{Medical Imaging 2019: Image Processing}, vol.
  10949.\hskip 1em plus 0.5em minus 0.4em\relax International Society for
  Optics and Photonics, 2019, p. 109491H.

\bibitem{alaverdyan2018unsupervised}
Z.~Alaverdyan, J.~Chai, and C.~Lartizien, ``Unsupervised feature learning for
  outlier detection with stacked convolutional autoencoders, siamese networks
  and wasserstein autoencoders: application to epilepsy detection,'' in
  \emph{Deep Learning in Medical Image Analysis and Multimodal Learning for
  Clinical Decision Support}.\hskip 1em plus 0.5em minus 0.4em\relax Springer,
  2018, pp. 210--217.

\bibitem{chen2019unsupervised}
X.~Chen, N.~Pawlowski, B.~Glocker, and E.~Konukoglu, ``Unsupervised lesion
  detection with locally gaussian approximation,'' in \emph{International
  Workshop on Machine Learning in Medical Imaging}.\hskip 1em plus 0.5em minus
  0.4em\relax Springer, 2019, pp. 355--363.

\bibitem{alaverdyan2020regularized}
Z.~Alaverdyan, J.~Jung, R.~Bouet, and C.~Lartizien, ``Regularized siamese
  neural network for unsupervised outlier detection on brain multiparametric
  magnetic resonance imaging: application to epilepsy lesion screening,''
  \emph{Medical image analysis}, vol.~60, p. 101618, 2020.

\bibitem{baur2021modeling}
C.~Baur, B.~Wiestler, M.~Muehlau, C.~Zimmer, N.~Navab, and S.~Albarqouni,
  ``Modeling healthy anatomy with artificial intelligence for unsupervised
  anomaly detection in brain mri,'' \emph{Radiology: Artificial Intelligence},
  vol.~3, no.~3, p. e190169, 2021.

\bibitem{heer2021ood}
M.~Heer, J.~Postels, X.~Chen, E.~Konukoglu, and S.~Albarqouni, ``The ood blind
  spot of unsupervised anomaly detection,'' in \emph{Medical Imaging with Deep
  Learning}, 2021.

\bibitem{baur2020steganomaly}
C.~Baur, R.~Graf, B.~Wiestler, S.~Albarqouni, and N.~Navab, ``Steganomaly:
  Inhibiting cyclegan steganography for unsupervised anomaly detection in brain
  mri,'' in \emph{International Conference on Medical Image Computing and
  Computer-Assisted Intervention}.\hskip 1em plus 0.5em minus 0.4em\relax
  Springer, 2020, pp. 718--727.

\bibitem{zimmerer2019unsupervised}
D.~Zimmerer, F.~Isensee, J.~Petersen, S.~Kohl, and K.~Maier-Hein,
  ``Unsupervised anomaly localization using variational auto-encoders,'' in
  \emph{International Conference on Medical Image Computing and
  Computer-Assisted Intervention}.\hskip 1em plus 0.5em minus 0.4em\relax
  Springer, 2019, pp. 289--297.

\bibitem{van2021anomaly}
K.~M. van Hespen, J.~J. Zwanenburg, J.~W. Dankbaar, M.~I. Geerlings,
  J.~Hendrikse, and H.~J. Kuijf, ``An anomaly detection approach to identify
  chronic brain infarcts on mri,'' \emph{Scientific Reports}, vol.~11, no.~1,
  pp. 1--10, 2021.

\bibitem{han2021madgan}
C.~Han, L.~Rundo, K.~Murao, T.~Noguchi, Y.~Shimahara, Z.~{\'A}. Milacski,
  S.~Koshino, E.~Sala, H.~Nakayama, and S.~Satoh, ``Madgan: unsupervised
  medical anomaly detection gan using multiple adjacent brain mri slice
  reconstruction,'' \emph{BMC bioinformatics}, vol.~22, no.~2, pp. 1--20, 2021.

\bibitem{you2019unsupervised}
S.~You, K.~C. Tezcan, X.~Chen, and E.~Konukoglu, ``Unsupervised lesion
  detection via image restoration with a normative prior,'' in
  \emph{International Conference on Medical Imaging with Deep Learning}.\hskip
  1em plus 0.5em minus 0.4em\relax PMLR, 2019, pp. 540--556.

\bibitem{zuo2018automatic}
L.~Zuo, A.~Carass, S.~Han, and J.~L. Prince, ``Automatic outlier detection
  using hidden markov model for cerebellar lobule segmentation,'' in
  \emph{Medical Imaging 2018: Biomedical Applications in Molecular, Structural,
  and Functional Imaging}, vol. 10578.\hskip 1em plus 0.5em minus 0.4em\relax
  International Society for Optics and Photonics, 2018, p. 105780D.

\bibitem{popescu2021distributional}
S.~G. Popescu, D.~J. Sharp, J.~H. Cole, K.~Kamnitsas, and B.~Glocker,
  ``Distributional gaussian process layers for outlier detection in image
  segmentation,'' in \emph{International Conference on Information Processing
  in Medical Imaging}.\hskip 1em plus 0.5em minus 0.4em\relax Springer, 2021,
  pp. 415--427.

\bibitem{bowles2017brain}
C.~Bowles, C.~Qin, R.~Guerrero, R.~Gunn, A.~Hammers, D.~A. Dickie, M.~V.
  Hern{\'a}ndez, J.~Wardlaw, and D.~Rueckert, ``Brain lesion segmentation
  through image synthesis and outlier detection,'' \emph{NeuroImage: Clinical},
  vol.~16, pp. 643--658, 2017.

\bibitem{zhang2020viral}
J.~Zhang, Y.~Xie, G.~Pang, Z.~Liao, J.~Verjans, W.~Li, Z.~Sun, J.~He, Y.~Li,
  C.~Shen \emph{et~al.}, ``Viral pneumonia screening on chest x-rays using
  confidence-aware anomaly detection,'' \emph{IEEE transactions on medical
  imaging}, vol.~40, no.~3, pp. 879--890, 2020.

\bibitem{ouardini2019towards}
K.~Ouardini, H.~Yang, B.~Unnikrishnan, M.~Romain, C.~Garcin, H.~Zenati, J.~P.
  Campbell, M.~F. Chiang, J.~Kalpathy-Cramer, V.~Chandrasekhar \emph{et~al.},
  ``Towards practical unsupervised anomaly detection on retinal images,'' in
  \emph{Domain Adaptation and Representation Transfer and Medical Image
  Learning with Less Labels and Imperfect Data}.\hskip 1em plus 0.5em minus
  0.4em\relax Springer, 2019, pp. 225--234.

\bibitem{quellec2016multiple}
G.~Quellec, M.~Lamard, M.~Cozic, G.~Coatrieux, and G.~Cazuguel,
  ``Multiple-instance learning for anomaly detection in digital mammography,''
  \emph{Ieee transactions on medical imaging}, vol.~35, no.~7, pp. 1604--1614,
  2016.

\bibitem{kuo2016framework}
C.-T. Kuo and I.~Davidson, ``A framework for outlier description using
  constraint programming,'' in \emph{Thirtieth AAAI Conference on Artificial
  Intelligence}, 2016.

\bibitem{bowles2016pseudo}
C.~Bowles, C.~Qin, C.~Ledig, R.~Guerrero, R.~Gunn, A.~Hammers, E.~Sakka, D.~A.
  Dickie, M.~V. Hern{\'a}ndez, N.~Royle \emph{et~al.}, ``Pseudo-healthy image
  synthesis for white matter lesion segmentation,'' in \emph{International
  Workshop on Simulation and Synthesis in Medical Imaging}.\hskip 1em plus
  0.5em minus 0.4em\relax Springer, 2016, pp. 87--96.

\bibitem{baur2021autoencoders}
C.~Baur, S.~Denner, B.~Wiestler, N.~Navab, and S.~Albarqouni, ``Autoencoders
  for unsupervised anomaly segmentation in brain mr images: a comparative
  study,'' \emph{Medical Image Analysis}, p. 101952, 2021.

\bibitem{kim2019deep}
M.~Kim, J.~Yun, Y.~Cho, K.~Shin, R.~Jang, H.-j. Bae, and N.~Kim, ``Deep
  learning in medical imaging,'' \emph{Neurospine}, vol.~16, no.~4, p. 657,
  2019.

\end{thebibliography}
\bibliographystyle{IEEEtran}

\end{document}